\documentclass[hasAbstract,wideAuthorBox]{csmagazine}

\DeclareUnicodeCharacter{1EBF}{\'{\^{e}}}

\usepackage{ulem}

\newcommand{\add}[1]{ {{#1}}}
\newcommand{\remove}[1]{ }
\usepackage{hyperref}

\title{Community Organizations: Changing the Culture in Which Research Software Is Developed and Sustained}

\author{%
	Daniel S. Katz\\ \ieeecsAffiliation{University of Illinois at Urbana-Champaign}\\
    Lois Curfman McInnes\\ \ieeecsAffiliation{Argonne National Laboratory}\\
	David E. Bernholdt\\ \ieeecsAffiliation{Oak Ridge National Laboratory}\\
    Abigail Cabunoc Mayes\\ \ieeecsAffiliation{Mozilla}\\
    Neil P. Chue Hong\\ \ieeecsAffiliation{Software Sustainability Institute, University of Edinburgh}\\
    Jonah Duckles\\ \ieeecsAffiliation{The Carpentries}\\ 
	Sandra Gesing\\ \ieeecsAffiliation{University of Notre Dame}\\
    Michael A. Heroux \\ \ieeecsAffiliation{Sandia National Laboratories}\\
	Simon Hettrick\\ \ieeecsAffiliation{Software Sustainability Institute, University of Southampton}\\
	Rafael C. Jimenez, \\ \ieeecsAffiliation{ELIXIR}\\
    Marlon Pierce\\ \ieeecsAffiliation{Indiana University}\\ 
	Belinda Weaver\\ \ieeecsAffiliation{The Carpentries}\\ 
	Nancy Wilkins-Diehr\\ \ieeecsAffiliation{University of California, San Diego}\\
}

\begin{document}


\ieeecsPageHeaders{THEME ARTICLE: Reusable Software}{Computing in Science \& Engineering}{Feature}

\ieeecsArticleTitle


\ieeecsInsertAuthor


\raggedright

\ieeecsAbstract{Software is the key crosscutting technology that enables advances in mathematics, computer science, and domain-specific science and engineering to achieve robust simulations and analysis for  science, engineering, and other research fields.  However, software itself has not traditionally received focused attention from research communities; rather, software has evolved organically and inconsistently, with its development largely as by-products of other initiatives.  Moreover, challenges in scientific software are expanding due to disruptive changes in computer hardware, increasing scale and complexity of data, and demands for more complex simulations involving multiphysics, multiscale modeling and outer-loop analysis.  In recent years, community members have established a range of grass-roots organizations and projects to address these growing technical and social challenges in software productivity, quality, reproducibility, and sustainability. This article provides an overview of such groups and discusses opportunities to leverage their synergistic activities while nurturing work toward emerging software ecosystems.}

\textit{\textcircled{c} 2018 IEEE, Computing in Science \& Engineering (CiSE), doi: \href{https://doi.org/10.1109/MCSE.2018.2883051}{10.1109/MCSE.2018.2883051}}
\RaggedRight




During the past twenty years, computation has penetrated essentially all areas of research, including science, engineering, technology, and society; advanced modeling, simulation, and data analysis drive new discoveries and new understanding as complements to experimental and theoretical methods.\cite{siam-cse18,GroppHarrisonEtAl2016}  Reusable software is a key element of these advances; software libraries and community codes encapsulate cutting-edge algorithms and domain-specific expertise, thereby enabling use/reuse and facilitating collaboration.
We can understand the impact of software in research across all fields by asking researchers, by examining their published papers, and by examining their funding.
Two surveys, of academic researchers at Russell Group universities in the UK~\citep{uk_survey} and members of the National Postdoctoral Association in the United States,\citep{nangia_survey} found that about 65\% of respondents said they couldn't do their research without software, while about 25\% said they could, but it would be much more difficult, and only a few percent said it would make no difference.
And a study of 40 papers in {\it Nature} from January to March 2016 showed that 32 explicitly mentioned software, with each paper mentioning an average of 6.5 software tools, almost all of which were research software.\citep{nangia_nature}
Likewise, searching the NSF award database for projects that mention ``software'' in their abstracts between 1995 and 2016 finds 18,592 awards totaling \$9.6 billion.\citep{dia2}

Software is essential to most of today's research, and the software used for a research project is almost never entirely developed by that project but, rather, it depends on, uses, and builds on research software from other projects and from other developers. Thus, it becomes apparent that research software, much like research itself, is actually developed and maintained by a community, forming an ecosystem of competing and collaborating products.  Much of this circumstance is due to the open source movement and its culture of sharing and collaboration, similar to the idealized culture of open science and open research toward which we are slowly moving.

Open source has created a tremendous variety of software, but this plethora of solutions is not easy for researchers to find and use out of the box.
Moreover, researchers face growing challenges in creating more ambitious software for all research areas.\cite{HerouxAllenEtAl2016, KeyesMcInnesWoodwardEtAl13} For example, in computational science and engineering,  challenges include coupling physics, scales, and analytics while adapting to disruptive changes in computing architectures.  Due to limitations in funding models and reward structures, researchers face pressure to publish new scholarly results quickly rather than investing in development of sustainable software that reliably supports longer-term research and interdisciplinary collaboration.  Researchers need training in best practices for software engineering, customized to address the unique needs of disciplinary cultures, yet typical graduate and undergraduate programs do not adequately cover these topics.
 
To address these circumstances and promote research collaboration through emerging software ecosystems, community members have recently established a variety of grass-roots organizations and projects, which have been further inspired by the growth of digital resources (that can more easily be shared), the growth of the internet (making sharing easier), and the growth of collaborative tools such as GitHub and Slack.
Community organizations that focus on a particular discipline, a particular technology, or particular functional skills can help researchers understand relevant parts of the ecosystem, including what software is available and what isn't and how the available software packages compare. In addition, these community organizations can support the health of the ecosystem, for example, by encouraging policies, reuse, and collaboration and supporting best practices for software development. The remainder of this paper highlights such organizations and how they work.


\section{Overview of community organizations}

In this section, we provide overviews of several community organizations: the Software Sustainability Institute (SSI), Conceptualization of a U.S. Research Software Sustainability Institute (URSSI), IDEAS Software Productivity, Better Scientific Software (BSSw), the Science Gateways Community Institute (SGCI), ELIXIR, Mozilla, the Apache Foundation, Software Carpentry, and WSSSPE. We also provide brief summaries of some organizations that have written longer articles in this special issue or are described in more detail elsewhere: Computational Infrastructure for Geodynamics (CIG), Molecular Science Software Institute (MolSSI), NuMFOCUS, rOpenSci, Extreme-scale Scientific Software Development Kit (xSDK), Astrophysics Source Code Library (ASCL), and an emerging effort on Promoting Research Software.  For each organization, we present its goals, including potential culture changes, some history, its activities, and how these are designed to achieve the organization's goals, and, when appropriate, information about the organization's future.

\subsection{The Software Sustainability Institute}

The Software Sustainability Institute (SSI, \url{https://www.software.ac.uk/}) was established in 2010 to cultivate better and more sustainable research software to enable world-class research: ``{\it Better software, better research}.'' The SSI is unique in supporting the mainstream of researchers across all disciplines. It has established itself as the de facto authority for research software practice, acting as a focal point to enable best practices to be shared within and between disciplines. 

The goal of the SSI is to enable the UK research community (and its international collaborators) to take full advantage of software and, in doing so, to support the conduct of excellent research. 

The SSI (a consortium of the universities of Edinburgh, Manchester, Oxford and Southampton) was formed after the conclusion of the UK's e-Science program brought a widespread call from the UK research community for the increased reliability and robustness of research software. Initial funding was from the UK Engineering and Physical Sciences Research Council, subsequently joined by the Economic and Social Sciences Research Council and Biotechnology and Biological Sciences Research Council, with the mandate to develop an understanding of the requirements and challenges of research software users and developers from all disciplines. To address these challenges, the SSI worked directly with over 80 groups to improve their codes and ran community engagement workshops in different domains. A fellowship program has created a network of over 100 advocates at 70 organizations from across research disciplines, types of institutions, and career stages. More recently, the SSI has transitioned to enabling communities to help themselves. This effort has included the collection of evidence and provision of arguments for changing stakeholder policy and engaging with new organizations and groups being set up in the UK and internationally aimed at supporting research software.

Clearly for any significant and permanent improvements in the practice of research software, structural problems around lack of skills and recognition must be addressed. The role of the SSI is to provide the support and leadership to create these cultural shifts.

The activities of the SSI~\citep{SSI} include
community engagement (annual Collaborations Workshop, topic and domain workshops, fellowship program, and support for other community events and initiatives); training (UK coordination of The Carpentries, instructor training, online guides); research software engineering (open call for consultancy projects, software evaluation service, software management plans); policy (campaigns and reports, research and data collection, contributing to international best practice); and outreach (blog and social media, {\it Journal of Open Research Software}).

The SSI believes that "communities of practice"~\citep{wenger} are the most effective and sustainable way of creating the cultural change that enables better practices to be widely adopted. The SSI's authority and support make a big difference to the success of these communities: taking an issue and turning it into a plan of action, leveraging its network of collaborators to bring the right people together, and empowering people to step up to leadership roles to take the community forward. For example,
the Research Software Engineer (RSE) movement started from an SSI workshop in 2012~\citep{rse_blog} and has led to a community of over 1,200 RSEs and chapters in Australia, Canada, Germany, the Netherlands, Norway, South Africa, and the United States. 
The Carpentries (see Software Carpentry, below) now number hundreds of instructors worldwide, but growth in the UK started with just one person when the SSI hosted the first UK training events in 2012. A focus on capacity growth---training 140+ instructors---now enables 1,600 researchers to be trained each year in the UK.

The SSI has developed an understanding of how research software is used and developed and how software is changing the way research is conducted.\add{A key lesson learned was to support, and collaborate with, enthusiastic leaders (e.g., Fellows, RSEs) who furthered the SSI's goals.} The SSI is now planning to focus on the creation of communities of practice to empower cultural change and enable far wider adoption of better practices, with the aim of sharing expertise, enabling software reuse, and providing support that scales to even greater numbers of people.

\subsection{Conceptualization of a U.S. Research Software Sustainability Institute}

An active NSF-funded project, Conceptualization of a U.S. Research Software Sustainability Institute (URSSI, \url{http://urssi.us}),\citep{urssi_cise} is currently working to make the case for and plan a possible institute to improve science and engineering research by supporting the development and sustainability of research software in the United States. The institute aims to address three primary classes of concern that are pervasive across research software in all research disciplines and have stymied research software from achieving maximum impact: functioning of the individual and team, functioning of the research software, and functioning of the research field itself.

The goal of this conceptualization project is to create a roadmap for a future URSSI (an actual institute) to minimize or at least decrease these types of concerns. In order to do so, the URSSI conceptualization has two aims: bring the research software community together to determine how to address the issues that are already known; and identify additional issues URSSI should address, identify communities for whom these issues are relevant, determine how to address the issues in coordination with the communities, and decide how to prioritize all the issues in URSSI.

URSSI is working with other community organizations, both outside and inside NSF. Many of these organizations are described in other papers in this special issue or in other subsections of this paper. In the NSF-funded space in particular, there are two ongoing institutes and a recently completed conceptualization, two other conceptualization projects now under way, and a large number of software development and maintenance projects. In the UK, the Software Sustainability Institute is an inspiration and a potential model for this project, as well as a potential collaborator. Figure 1 shows a view of the space of URSSI, which is similar to that of the SSI.  Where a disciplinary or technology project/community exists, URSSI will work with it, taking successful activities and generalizing them for other communities.  And where there is no organized community, URSSI will work directly with software stakeholders.  

\begin{figure}[h]
	\begin{center}	
		\includegraphics[width=0.5\textwidth]{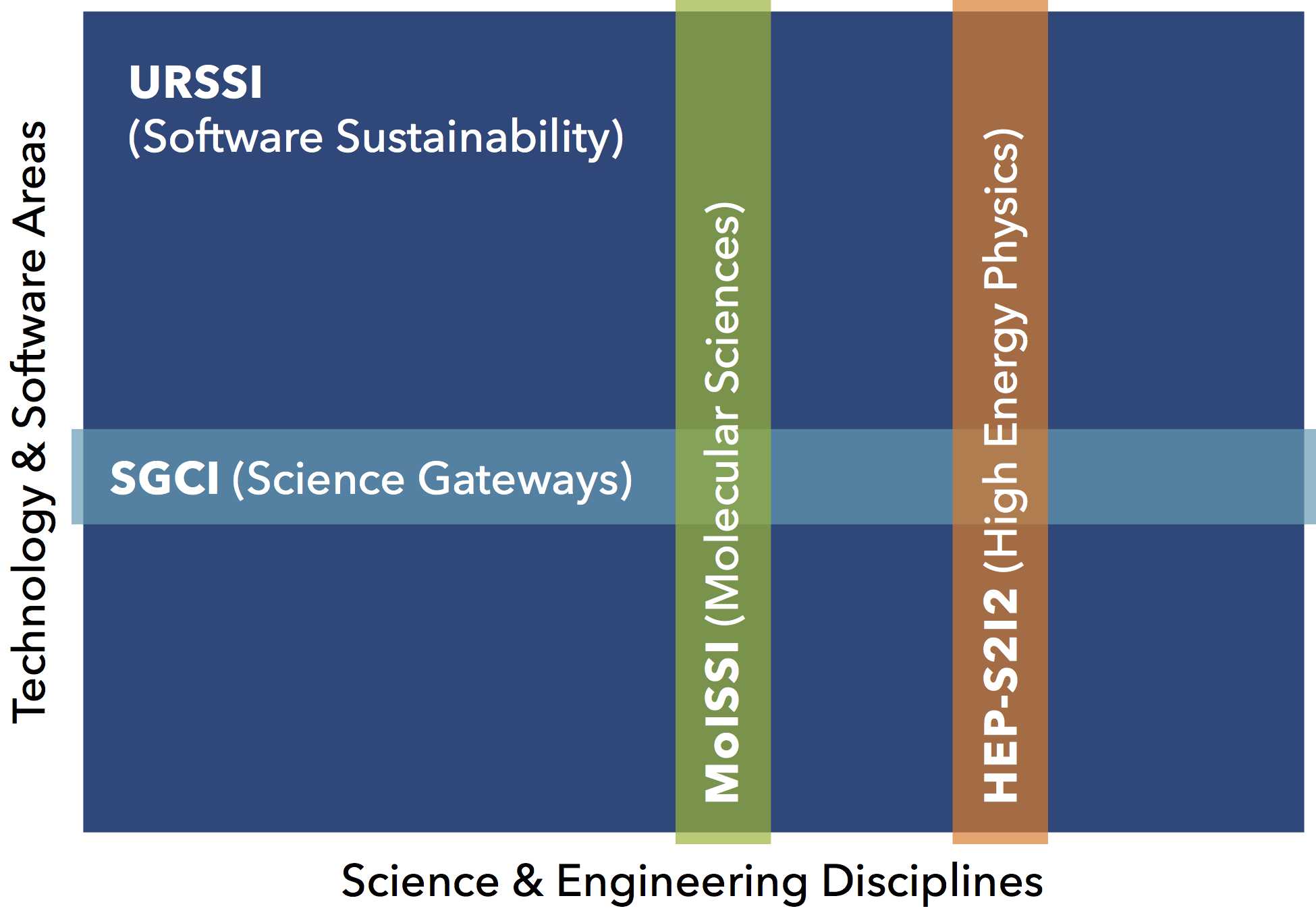}
		\caption{General space in which a U.S. research software sustainability institute would operate, showing overlaps with other NSF-funded software institutes. \label{fig:urssi}}		
	\end{center}
\end{figure}

Given these existing organizations, part of the challenge is to define how URSSI will work with these other groups. For example, URSSI might decide that another group performs an activity so well that URSSI should point to it, such as the SSI's software guides. Or URSSI might decide either to duplicate or to enhance an activity another group does, in order to expand its impact, such as working with the Science Gateway Community Institute to offer incubator services to a wider community than just gateway developers. Or URSSI might decide to collaborate with one or more groups, such as on policy campaigns aimed at providing better career paths for research software developers in universities.

URSSI itself is working through a series of activities, including detailed ethnographic studies of three specific projects, two focused workshops with 20--30 participants on software credit and software project incubation, two more general workshops with 50--80 participants aimed at discussion of challenges and potential solutions, and a widely distributed community survey. In addition, to bring the U.S. research software community together around URSSI, the project has a website, a series of newsletters, a set of URSSI-written and community-contributed blogs, and a community discussion site.

The immediate goal of the URSSI conceptualization project is to make the case for and plan a possible institute to improve science and engineering research by supporting the development and sustainability of research software in the United States.  Making the case involves bringing the research software community together around the idea of an institute, and both the project and the community convincing NSF and/or another funding agency to offer an opportunity to run an institute.  The institute would then work with the community to improve the processes that are used and the policies that shape how research software is developed, used, and maintained.

\subsection{IDEAS Software Productivity} 

Researchers in computational science and engineering (CSE) face unprecedented challenges due to a confluence of disruptive changes in HPC architectures and opportunities for next-generation simulation, analysis, and design. Teams are working toward predictive science through multiphysics, multiscale simulations and analytics, while needing greater scientific reproducibility and exploiting massive on-node concurrency during disruption in underlying hardware, system software, and programming environments. This situation brings with it an opportunity to fundamentally change how scientific software is designed, developed, and sustained.  The Interoperable Design of Extreme-scale Application Software project (IDEAS, \url{https://www.ideas-productivity.org}) is partnering with the CSE community to improve developer productivity (positively impacting software quality, development time, and staffing resources) and software sustainability (reducing the cost of sustaining and evolving software over its intended lifetime)---thereby helping improve scientific productivity while ensuring continued scientific success.  

The IDEAS project began in 2014, sponsored by the U.S. Department of Energy Office of Science, to address challenges in software productivity and sustainability, with emphasis on terrestrial ecosystem modeling.  The project expanded in 2017 in the DOE’s Exascale Computing Project (ECP, \url{https://www.exascaleproject.org}), which requires intensive development of applications and software technologies while anticipating and adapting to continuous advances in computing architectures. The role of IDEAS within the ECP is to ease the challenges of software development and ensure that investment in the exascale software ecosystem is as productive and sustainable as possible.

A central IDEAS activity is productivity and sustainability improvement planning (PSIP),\cite{psip_tools} a lightweight, iterative workflow where teams identify their most urgent software bottlenecks and work to overcome them.  To support these advances, the IDEAS project collaborates with the community to develop, customize, and curate software methodologies, processes, and tools for improving software productivity and sustainability.  Topics include software engineering, design, development, testing, refactoring, and performance, with information tailored to address the needs of high-performance CSE.  Resources are relevant for projects of all sizes, ranging from small teams (for example, a professor and students) to aggregate teams (that is, multiple teams who collaborate on next-generation science via software yet must maintain autonomy and flexibility because of varying priorities, stakeholders, and funding).  

To serve and partner with the broader community, IDEAS outreach~\citep{bernholdt_ideas_outreach_blog2018} features a webinar series on {\it Best Practices for HPC Software Developers}, various tutorials and events,\citep{ideas_events_website} and content from the Better Scientific Software site (see below).  Complementary work focuses on building an extreme-scale scientific software ecosystem composed of high-quality, reusable software components and libraries (xSDK; see below).  

IDEAS is working to advance the quality of high-performance CSE and to provide a foundation (through software productivity methodologies and an extreme-scale software ecosystem) that enables transformative and reliable next-generation predictive science and decision support.  
Members of the IDEAS project are pursuing collaboration with synergistic groups to work toward long-term changes in the culture, funding, and reward structure of CSE. 
\add{A key lesson from IDEAS has been the value in using webinars to help to engage a broader audience and to get the project's message out.}

\subsection{Better Scientific Software}

The Better Scientific Software site (\url{https://bssw.io}) is a new community-based resource for scientific software improvement---a central hub for sharing information on practices, techniques, experiences, and tools to improve developer productivity and software sustainability for CSE and related technical computing areas. The site features curated content, experiences, and reasoned insights provided by the international community, including researchers, practitioners, and stakeholders from national laboratories, academia, and industry who are dedicated to curating, creating, and disseminating information that leads to improved CSE software.   Historically, opportunities for CSE software developers to exchange information and experiences have been limited; BSSw provides a space to support this kind of sharing.

The BSSw site was launched in November 2017 as an outgrowth of the IDEAS project (see above).  The long-term vision for BSSw is to serve as an international community-driven and community-managed resource, with content and editorial processes provided by volunteers, initially nucleated by the IDEAS team but over time expanding to much broader participation.  The BSSw Fellowship program gives recognition and funding to leaders and advocates of high-quality scientific software. 

The BSSw platform provides easy access to resources and training materials provided by the community. Content spans a range of topics (such as scientific software planning, development, reliability, collaboration, and performance), including introductory {\it WhatIs} and {\it HowTo} information covering basic steps for improving software productivity and sustainability.  The site includes a growing collection of curated content---brief articles that highlight other web-based materials, describing why the scientific software community might find them of value.  The site also features an expanding collection of original blog articles, including science teams’ experiences with productivity-related software issues and discussion of organizations in the software productivity community. 

The BSSw site is the starting point for any new user. A GitHub backend enables content development using a collaborative, open workflow.  Content can also be contributed with an easy-to-use Google form.  Anyone with experience or expertise who can help other scientific software teams is encouraged to contribute an article or pointer to relevant work.\cite{bssw_contributing}

BSSw encompasses a rich variety of communities who are working to advance the methods and practices of CSE software. BSSw community landing pages provide custom starting points for using the site and promote shared understanding of scientific software issues.  Curators of a community landing page can customize content to serve the needs of community members.  BSSw communities include a growing set of science-focused areas (computational molecular sciences, environmental system science) as well as crosscutting areas (scientific libraries, software engineering, supercomputing facilities and their users, and exascale computing).

By providing a venue to share information and experiences on software issues, BSSw is raising awareness of the importance of good software practices to scientific productivity and enabling readers to discover potential connections to their own priorities and workflows.  Future plans center on establishing broader community leadership and growth in order to help CSE researchers (regardless of nationality and funding sources) to increase software productivity, quality, and sustainability while changing CSE culture to fully support software's essential role.
\add{BSSw has shown that it is possible to lower the barrier to participation by leveraging an environment with which the majority of the community is already familiar (GitHub and Markdown).}

\subsection{Science Gateways Community Institute}

The Science Gateways Community Institute (SGCI, \url{https://sciencegateways.org}) was launched in 2016 in recognition of the important role that end-to-end solutions, also called science gateways, play in the advancement of research. Often, gateways are the means by which millions of dollars of research infrastructure---telescopes and microscopes, supercomputers and sensor streams---are accessed. All gateways typically have an intuitive user interface component. Many are completely open, democratizing access to high-end resources typically available only to those at leading research institutions.
 
But the designers of SGCI had observed that gateway developers did not interact with one another. The goal of SGCI is to help the research community build more cost-effective, sustainable science gateways. Seven years of focus group studies and a large-scale survey informed the design of the Science Gateways Community Institute~\citep{SGCI-cise} and its five service areas: Incubator, Extended Developer Support, Scientific Software Collaborative, Community Engagement and Exchange, and Workforce Development.
 
The Incubator functions much like a business incubator, providing specialized expertise for short durations of time, and also runs a one-week bootcamp for gateway teams. Extended Developer Support provides up to 12 months of hands-on assistance in actually developing a gateway, including technology selection and best practices such as version control, open source, and support for community contributions. The Scientific Software Collaborative team develops and maintains a catalog of functional science gateways available worldwide, including links to the software packages used to build these gateways. Community Engagement and Exchange (CEE) fosters the gateway community, primarily through an annual conference with citable proceedings and the opportunity for authors to publish in a special issue of a peer-reviewed journal jointly managed with gateway organizations in Europe and Australia. Additional CEE activities include webinars, newsletters, and website content that engage the community. Workforce Development encompasses SGCI’s student programs, with several components such as coding institutes, internship placements with staff and clients, hackathons, and travel support to attend gateway-related events.
 
Through these activities, SGCI endeavors to create an international culture and a local support structure that lead to the development of more sustainable, effective gateways that advance science. SGCI's long-term goal is to distribute gateway-building expertise throughout that community. This is envisioned in several ways.\citep{SGCI-road, SGCI-2years} Because most academic development of science gateways happens on campuses, changing the landscape there is important. Several institutions are part of SGCI specifically because they are models of how successful gateway developer groups can be set up on campus. SGCI has engaged with many university campuses as well as groups such as the XSEDE Campus Champions (\url{https://www.xsede.org/community-engagement/campus-champions}) and ACI-REF (\url{https://aciref.org}) to share and promote these solutions and support career paths for developers and research computing engineers.

\remove{Future directions for other areas of SGCI include a focus on scalability. For example, cohorts graduating from SGCI's bootcamp can continue to collaborate.}

\add{Over the first two years, one of the lessons the SGCI team has had to learn quickly is how to scale due to demand. This scaling is envisioned throughout the organization. Cohorts graduating from SGCI's bootcamp can continue to collaborate through the formation of a brain trust.}
EDS activities can be made more scalable by sharing case studies and straightforward gateway development solutions that may meet the needs of many, without a long-term engagement with SGCI. Trained students will benefit others as they take their science gateway knowledge gained through SGCI internships and share it with others over their careers. SGCI envisions these mechanisms causing a positive change in how science gateway development is approached.

\subsection{ELIXIR}

ELIXIR (\url{https://www.elixir-europe.org}) is an intergovernmental organization that brings together life science resources across Europe. These resources include databases, software tools, training materials, cloud storage, and supercomputers. The goal of ELIXIR is to coordinate these resources so that they form a single infrastructure, making it easier for scientists to find and share data, exchange expertise, and agree on best practices.

The ELIXIR Scientific Advisory Board, supported by the ELIXIR preparatory phase reports, emphasized the importance of ensuring the quality and sustainability of software developed in ELIXIR. To support this goal, at the end of 2015, ELIXIR created the ``Software development best practices'' group in partnership with the Software Sustainability Institute and the Netherlands eScience Center. The group includes experts from ELIXIR nodes as well as other stakeholders concerned with the quality and sustainability of software for research across scientific disciplines.

The group began by defining good software development practices as well as metrics to assess their adoption.\citep{artaza} Later the group proposed the 4OSS recommendations, a smaller set of broader recommendations to support the understanding and uptake of these practices.\citep{jimenez} The 4OSS recommendations are designed around open source values and provide practical suggestions that contribute to making research software and its source code more discoverable, reusable, and transparent. The recommendations are meant not just for software developers but also for research funders, research institutions, journals, group leaders, and managers of projects producing research software.

The current activities of the group focus on assessing the adoption of the 4OSS recommendations and training to facilitate their adoption. The ELIXIR benchmarking framework OpenEbench (\url{https://openebench.bsc.es}) is being used to assess compliance; and \url{https://bio.tools}, the ELIXIR registry of tools,\citep{ison} to visualize that compliance. In partnership with the ELIXIR training platform, Software Carpentry, and other communities, the group is also creating a collection of training materials to help researchers and developers implement the 4OSS recommendations. ELIXIR is also exploring how to improve the quality of their software in terms of user experience and usability. All these activities aim to trigger the development of higher-quality and more-sustainable software, with the ultimate goal of helping researchers perform better research.

\add{A key lesson learned by ELIXIR has been to focus on a simple, practical, and achievable set of outcomes, engaging with communities beyond ELIXIR. These outcomes should be able to be easily adopted and measured by the community, linked to other community efforts and supported by a wide range of stakeholders.}

\subsection{Mozilla} 

The mission of Mozilla (\url{https://www.mozilla.org}) is to ensure the internet is a global public resource, open and accessible to all. Its global network of technologists, thinkers, and builders work for a healthier internet through core issues such as privacy and security, openness, decentralization, web literacy, and digital inclusion. Among the research community, Mozilla supports project-based learning and mentorship to further data sharing and open source, provides fellowships and mini-grants to empower the next generation of leaders, and advocates for the growing number of researchers working openly.

Mozilla's strategy is to fuel a movement that connects a global force of people willing to stand up for and build the open internet. Researchers play a major role since scientists invented the web and research increasingly relies on the open source software that also drives the open internet. By investing in open science leaders, Mozilla wants to see breakthroughs that bring together researchers in this movement.

Since 2013, Mozilla has worked with the scientific community to make research more accessible, open, and interoperable. Open source is the backbone of scientific software today. Adopting open source software practices leads to more sustainable and innovative research software. With roots in the open source movement, Mozilla continues to work openly, train others, and research open source software.

Mozilla has three main activities addressing the culture around scientific software:
\begin{itemize}
\itemsep0em 
\item Fellowships. The Mozilla Fellowships for Science (\url{https://science.mozilla.org/programs/fellowships}) promote open science, open access, and open source in research programs globally. Fellows are funded for 10 months of research in their local institutions and scientific departments.

\item Mini-Grants. The Open Science Mini-Grants (\url{https://science.mozilla.org/blog/2018b_minigrantrfp}) fund project proposals focusing on prototyping, community building, or curriculum. These projects work toward open innovation, efficiency in open science, and reproducibility.

\item Open leadership training. Mozilla Open Leaders (\url{https://foundation.mozilla.org/opportunity/mozilla-open-leaders/}) provides mentorship and training on working open best practices. Participants join a cohort of open leaders for 14 weeks to design and build projects that empower others to collaborate within inclusive communities.
\end{itemize}

The Mozilla Fellowships for Science empower open science advocates embedded in academic institutions. Fellows receive training and support on open source, data sharing, open science policy, and licensing. They act as local catalysts by crafting code and learning resources that help their communities adopt open practices, and they teach their institutional peers.
Through the Open Science Mini-Grants, Mozilla supports and develops a community of leaders with the aim of transforming research and the culture around science.
Mozilla Open Leaders provide the training and community support needed to build an open project through cohort-based training and 1:1 mentorship. Graduates are encouraged to return as mentors, thus transforming participants into mentors and advocates, helping the next generation work openly.

Mozilla will continue to connect leaders motivated by this agenda to bring the next wave of openness and opportunity in online life. In turn, this activity will make research more open, accessible and interoperable.

\subsection{Apache Software Foundation}

The Apache Software Foundation (ASF, \url{https://apache.org}) is a 501(c)3 nonprofit organization that fosters the growth of open source software communities and provides technical infrastructure and support mechanisms, including licensing and legal support, that are needed by these communities.  ASF can be thought of as a ``factory'' for open source communities, providing flexible templates for organization and governance.  In 2018, the foundation consisted of 322 open source projects, as well as 52 incubating projects that are in the process of joining the foundation. 

Although the ASF typically does not work directly with communities associated with scientific code bases and many of its member projects have their roots in commercially produced software, the roots of the foundation itself are based on academically produced software (the HTTPD server) that had its origin in supporting the international high-energy physics community. 

The ASF's founding goals (\url{https://www.apache.org/foundation/how-it-works.html}) are fourfold:
\begin{itemize}
\itemsep0em 
\item Provide a foundation for open, collaborative software development projects by supplying hardware, communication, and business infrastructure;
\item Create an independent legal entity to which companies and individuals can donate resources and be assured that those resources will be used for the public benefit; 
\item Provide a means for individual volunteers to be sheltered from legal suits directed at the foundation's projects; 
\item Protect the "Apache" brand, as applied to its software products, from being abused by other organizations.
\end{itemize}

Founded in 1999, the foundation grew from a group of maintainers of the open source HTTPD server code initially developed by the National Center for Supercomputing Applications.  
The distributed locations and diverse employments of the founding members greatly shaped the organization's focus on meritocratic decision making processes, legal cover for individuals (ensuring members have freedom to act independently of their employers' intellectual property and copyright policies), and licensing approach. In particular, the foundation's license allows proprietary extensions (unlike the ``free'' licenses of the Free Software Foundation) and thus is potentially more amenable as a basis for commercial software. 

As an umbrella organization for its constituent projects, ASF has three primary activities: operating the Incubator, a metaproject that helps projects join the foundation and understand its governance processes; operating the INFRA project, which provides infrastructure support (including mailing lists, code repositories, and issue-tracking systems); and conducting two annual ApacheCon conferences (one in North America and one in Europe) that promote member projects' software and enable collaboration among project members. 

A key cultural feature of the ASF is its focus on meritocracy and consensus-based governance with transparent, archived communications and decision-making mechanisms.  A common ASF slogan is ``community over code,'' meaning that a large, diverse, and vibrant developer base is the best chance that an open source project has for long-term sustainability. 

\subsection{Software Carpentry}

Software Carpentry (\url{https://software-carpentry.org}) is a community-driven project teaching foundational coding skills to researchers, empowering them to develop research software, automate research tasks and workflows, and perform reproducible science. Software Carpentry's approach has been to develop a volunteer community of trained peer instructors (often researchers themselves) who can capably teach other researchers. Software Carpentry workshops deliver community-developed, open source lessons on software and coding skills in an engaging workshop environment with a strongly enforced code of conduct.

Software Carpentry's more than 2,000 certified instructors are drawn from the research and research support communities and have completed a two-day instructor training course. In that course, they learn sound, evidence-based pedagogical practices that will make the sharing of their knowledge, skills, and abilities more impactful and engaging. Together with prepared instructors and community-developed lessons, Carpentries workshops have good outcomes (\url{https://carpentries.org/assessment}). Workshop attendees are taught to adopt a range of practices---version control, testing, automation---that foster open science and reproducibility, deliver greater efficiencies, reduce the risk of error, and build confidence in further learning.

Software Carpentry had its start in 1998 teaching good coding practices to researchers at Los Alamos National Laboratory. In 2012, with support from the Alfred P. Sloan Foundation, Software Carpentry pivoted to its instructor training and capacity-building model. The launch of the two-day workshop model accelerated its growth and uptake. In 2015 an annual membership model was launched to enable organizations to train local instructors and support the infrastructure provided by Software and Data Carpentry together. By mid-2018, The Carpentries (\url{https://carpentries.org}), a fiscally sponsored project of community initiatives, had more than 70 member organizations in 10 countries and an instructor community of 2,000 trained instructors in 39 countries.

Key activities of Software Carpentry include two-day workshops; instructor training by certified instructor trainers; collaborative, open lesson development; a mentoring program; and support for ongoing community building at the local level through the Champions network and the Community Cookbook (\url{https://cookbook.carpentries.org}). Software Carpentry also provides supporting infrastructure to allow community members to register workshops, apply for instructor training, run assessment surveys, and engage in targeted discussions via a range of mailing lists. Software Carpentry publishes a newsletter every two weeks. Other lesson communities within the Carpentries---Data Carpentry (\url{https://datacarpentry.org}), focused on reproducible data analysis workflows, and Library Carpentry (\url{https://librarycarpentry.org}), focused on the work of 21st-century research librarians---are extending the model of Software Carpentry to new areas.

Key outcomes are building skills and confidence in coding for research and in the teaching of peers with an aim to empower the participating community. Community members are the key drivers of a range of local community and skill-building activities such as study groups, hacky hours, skill-shares, research bazaars, hackathons and do-a-thons. The Carpentries encourage a growth mindset among digitally skilled researchers and help people embrace a culture of lifelong learning in community. Maintainers develop transferable skills in software development and managing collaborative projects, while instructors develop research and teaching insights by instructing learners in other disciplines and gain useful perspectives and adaptability through teaching audiences with varying levels of skill. 
\add{Overall, Software Carpentry has demonstrated that community, shared lesson development for commonly taught topics, and a focus on improving teaching skills among the research community can make technical skills workshops more impactful and empowering for attendees.}

\subsection{Working towards Sustainable Software for Science: Practice and Experiences}

WSSSPE (\url{http://wssspe.researchcomputing.org.uk}) is an international community-driven organization that promotes sustainable research software by addressing challenges related to the full lifecycle of research software through shared learning and community action.  WSSSPE envisions a world where research software is accessible, robust, sustained, and recognized as a scholarly research product critical to the advancement of knowledge, learning, and discovery.

WSSSPE promotes sustainable research software by positively impacting the following areas:
\begin{itemize}
\itemsep0em 
\item Principles and best practices. Promoting best practices in sustainable software
\item Careers. Developing and supporting career paths in research software development and engineering
\item Learning. Engaging in activities to promote peer learning and interaction
\item Credit. Ensuring recognition of research software as an intellectual contribution equal to other research products
\end{itemize}

WSSSPE started with a one-day workshop at the SC13 conference in 2013 and gradually became a community.  In 2015 and 2016, WSSSPE used longer multiday meetings to attempt to create ongoing working groups but discovered that people who were willing to come and work together during a workshop could not necessarily also commit to doing large amounts of work outside those workshop days.  \add{A lesson learned from WSSSPE was that some}\remove{Some}working groups were successful, particularly those that overlapped already existing activities, but those that attempted to start new activities were generally not successful.

Recognizing this situation, WSSSPE has moved to become a series of one-day meetings associated with other events (in 2017 with the RSE conference and with the IEEE International Conference on eScience, and in 2018 with the IEEE International Conference on eScience again).  Thus, members of the WSSSPE community come together, discuss progress on ongoing activities, and bring in new members to such activities but generally do not create new activities solely under the WSSSPE umbrella.

\subsection{Additional Organizations}

In addition to the organizations discussed previously in this article, longer descriptions of some community organizations appear in this\add{pair of} special \remove{issue} \add{issues}, and descriptions of other organizations have appeared elsewhere. We briefly summarize several such organizations here, along the same lines (goals, history, activities, and future) as for the previously described organizations.

\subsubsection{Computational Infrastructure for Geodynamics}


Computational Infrastructure for Geodynamics~\citep{cig-here} (CIG, \url{https://geodynamics.org}) is a "community of practice" that advances Earth science by developing and disseminating software for geophysics and related fields.
CIG began in 2005 with the goal of advancing solid-Earth science and related fields by developing and disseminating scientific software, using best practices from computational science.
CIG identifies and encourages use of the best practices that are have proven to be most effective for this community, while also supporting development and dissemination of high-quality, free, open source scientific software for geophysics. CIG helps software authors improve their development practices and teaches early career scientists in particular to work with and extend these software packages, to prepare the scientific workforce to be expert users, and to contribute to scientific software.
Software is developed differently today in the geodynamics community from how it was in 2005, thanks to this organization. The CIG best practices, including use of open source licenses, version control, and automated test suites, have become the accepted standard in the community. CIG continues to improve these best practices and teach them to the community.

\subsubsection{The Molecular Science Software Institute}

The Molecular Science Software Institute (MolSSI, \url{https://molssi.org}) is a nexus for science, education, and cooperation serving the worldwide community of computational molecular scientists---a broad field including biomolecular simulation, quantum chemistry, and materials science. MolSSI began at the same time as the SGCI and went through a similar community-building and requirements-gathering process. MolSSI includes a software engineering team who develop software, interact with community software developers, support forums for standards development, mentor MolSSI software fellows, and work with industrial, national lab, and international partners.  The MolSSI software fellows are graduate students and postdocs, funded for up to two years, who work on their own software projects as well as outreach and education; they form a link between the institute and the larger community. MolSSI also provides coordination with the developers of community codes to improve their interaction with one another.

\subsubsection{NumFOCUS}

NumFOCUS (\url{https://numfocus.org}) is a 501(c)(3) nonprofit promoting open code for better science. The mission of NumFOCUS is to promote sustainable high-level programming languages, open code development, and reproducible scientific research. It accomplishes this mission through its educational programs and events as well as through fiscal sponsorship of open source scientific computing projects. NumFOCUS aims to increase collaboration and communication within the data science and scientific computing community. It was founded in 2012 with a goal to advance the long-term sustainability of open source scientific computing projects, working with both research and industry partners. Today NumFOCUS supports 23 fiscally sponsored projects and 23 affiliated projects, including popular data science tools such as Jupyter, pandas, and NumPy. It also organizes the PyData network, an educational program of regional events and local chapters spanning 45 countries and comprising over 85,000 data enthusiasts and practitioners. NumFOCUS works to connect the large global community of users of open source scientific software with the developers and maintainers of these projects in an effort to produce a supportive, sustainable future for these tools.

\subsubsection{rOpenSci}

The rOpenSci (https://ropensci.org) project~\citep{ropensci} is a nonprofit initiative founded in 2011 to enable open and reproducible research by creating technical infrastructure in the form of carefully vetted, staff- and community-contributed R packages that lower barriers to working with scientific data sources on the web; creating social infrastructure through a welcoming and diverse community; making data, tools and best practices more discoverable; building capacity of software users and developers; and advocating for a culture of data sharing and reusable software, all run by  a staff of research software engineers and a community manager.
rOpenSci has developed an ecosystem of hundreds of free and open source R tools. It runs a transparent, non-adversarial system for peer review of these packages, where successful authors can blog about their package to an audience of users and potential contributors. rOpenSci also provides a newsletter, blog posts, tweets using the \#rstats hashtag, a dedicated public discussion forum, and a Slack workspace for rOpenSci contributors.
rOpenSci hosts an annual hackathon-flavored "unconference" that brings together 60 current and new community members (users and contributors) to collaborate on projects they choose. The unconference integrates new members into the community, building an in-person trust network that contributes to the sustainability of their global online community. 

\subsubsection{Extreme-scale Scientific Software Development Kit}

Work on the xSDK (\url{https://xsdk.info}) began in 2014 as part of the IDEAS Software Productivity project (see above) in recognition that collaboration across independent numerical library efforts could have a tremendous positive impact on  developer productivity, software sustainability, and the capabilities that the libraries provide to users. Work focuses on community development and a commitment to combined success via quality improvement policies, better build infrastructure, and the ability to use diverse, independently developed libraries in combination to solve large-scale multiphysics and multiscale problems.  xSDK community policies (\url{https://xsdk.info.policies}) govern activities and set expectations for future xSDK members. Any package that satisfies the community policies is welcome. The first xSDK release in April 2016 included four widely used numerical libraries and a biogeochemistry package. In 2017, xSDK transitioned to become a primary delivery mechanism for math library capabilities in the Exascale Computing Project, incorporating continual advancements toward support for predictive science. The xSDK release in December 2017 included a total of 9 packages.  Eleven additional packages are working toward inclusion the fall 2018 xSDK release (for a total of 20 xSDK packages). The long-term goal of xSDK is community collaboration toward productive and sustainable community software ecosystems.

\subsubsection{Astrophysics Source Code Library}

The Astrophysics Source Code Library (ASCL, \url{https://ascl.net}) seeks to improve the transparency and reproducibility of astronomy and astrophysics by making the software enabling research results more discoverable for examination. The ASCL effort was founded in 1999; its editors seek out both new and old peer-reviewed papers that describe methods or experiments that involve the development or use of source code, and they add entries for the found software to the library if the source code is available for download. This approach ensures that source codes are added without requiring authors to actively submit them, resulting in a comprehensive listing that covers a significant number of the astrophysics source codes used in peer-reviewed studies. To change the culture of the astronomy community, the ASCL organization also encourages software authors to release and submit their own codes; and it works with authors, journals, and indexers to improve recognition of those who write software. The ASCL organization provides a citable software ID that is widely accepted by journals and indexers, advocates for citation of software on its own merit and career opportunities for software authors through participation in various groups and conference presentations, organizes conference sessions to improve the visibility of software authors, and links codes with the research literature it enabled.

\subsubsection{Promoting Research Software} 

Michelle Barker of the Australian Research Data Commons, working with Daniel S. Katz and Neil Chue Hong, has recently begun a "community of communities" activity informally being called Promoting Research Software. This community will have as a goal bringing together the various research software communities, many of which are elsewhere discussed in the article, to collectively promote research software as a critical research enabler.  This activity began with a small meeting in March 2018 and held additional meetings in September and October, alongside other events.  This metacommunity is planning to help the set of communities interested in research software develop a common message and work collaboratively toward larger goals, such as increased representation of research software in international initiatives or forums.



\section{Conclusions}



The work of these community organizations focuses on a few common themes, with variations in approaches according to  factors such as topical scope and sources of funding.
\add{While each organization focuses on a small number of activities to have impact, these choices are different for each organization. In a vast space like research software development, there are so many useful activities that could help, and there is a temptation to try to cover all the bases. But each successful organization has enough discipline not to try to take on too much too quickly. For example, the SSI is intentionally transitioning from one focus to a new one.}

In their respective spheres of influence, these groups nurture communities, work to change research culture, and promote the growth of software ecosystems.
Groups provide information about effective approaches for creating, sustaining, and collaborating via scientific research software.
While some groups establish partnerships with research software teams to provide guidance and intensive hands-on assistance for improved software practices, this approach has scalability limits in terms of human effort and cost. Other groups instead create flexible resources for self-study and outreach to the broader scientific research community.
Some groups provide modest funding and community recognition to promote attention to software quality improvements and to nurture new generations of leadership.
Common activities include establishing events (such as workshops, conferences, minisymposia, webinars, and tutorials) and informal mechanisms for sharing information (blog series and websites with community input).
Groups also articulate key issues and needs to stakeholders, agencies, and the broader research community, while advancing understanding of the importance of good-quality software to the integrity of computational research and to effect changes in policies, funding, and reward structure.
\add{As these organizations have grown, they have determined geographic scaling models that allow for coordination of contribution and effort.  Whether scaling to global, national, or regional impact, it has been important for these organizations to focus initially on empowering participants at the local scale. In addition, each organization's macro-level vision for shared impact has been critical. This approach of top-down vision meeting bottom-up local activities for developing impact is common across the successful organizations in this paper. Overall, these organizations seek impact and have extensive outreach activities; they are not merely trying to grow their communities but also are exploring ways to have impact outside their immediate communities.}

Important differences among these organizations also exist. Some differences are inherent to their scope and goals, such as SGCI aiming to benefit the science gateway community vs.\ ELIXIR working in life sciences. But others are related to how the organizations started and how they operate (governance). Some organizations started as broad communities, without dedicated funding, such as Apache, Software Carpentry, and WSSSPE. Others began as funded projects, aimed at a particular goal and then later expanding to try to change a community, perhaps because it became apparent that culture change was a key method for achieving the initial goal. Most of the organizations are led by participants who are now funded to be the leaders.  One issue is how this structure influences the sustainability of the organization itself.  When some participants in an organization are funded and others are not, conflicts can arise.  And if a transparent method does not exist for unfunded contributors to become funded and to become leaders, these unfunded contributors may not continue.  Most of the organizations presented here are technically projects, but some are fully independent (such as Apache and Mozilla).

While these community groups already interact and in some cases collaborate, opportunities abound for more focused partnerships to promote awareness of resources and fully leverage each group’s unique capabilities and outreach materials.  For example, almost all the organizations share an explicit or implicit long-term goal of changing the culture of research software and in turn enabling software to fully realize its role as a cornerstone of long-term collaboration and scholarly progress.
The Promoting Research Software activity aims at taking a first step toward this goal.  Many other areas of joint interest exist, where additional coordination and collaboration would serve the overall research community.



\ieeecsAcknowledgmentsHeader{ACKNOWLEDGMENTS}

\begin{ieeecsAcknowledgment}The
authors acknowledge the contribution of Alice Allen to this paper. The
Software Sustainability Institute acknowledges the support of the EPSRC through Grant EP/H043160/1 and EPSRC, ESRC and BBSRC through Grant EP/N006410/1.
URSSI is supported by the National Science Foundation (NSF) award 1743188.
IDEAS, BSSw, and xSDK acknowledge support from the Exascale Computing Project (17-SC-20-SC), a collaborative effort of the U.S. Department of Energy Office of Science and the National Nuclear Security Administration.
SGCI is supported by NSF award 1547611.
ELIXIR acknowledges the support of the ELIXIR-EXCELERATE grant funded by the European Commission within the Research Infrastructures program of Horizon 2020, grant agreement number 676559.
Mozilla Science is supported by the Alfred P. Sloan Foundation, the Helmsley Charitable Trust, and the Siegel Family Endowment.
The Carpentries is supported by the Alfred P. Sloan Foundation, the Gordon and Betty Moore Foundation, and the ongoing annual membership contributions of more than 70 member research organizations in 10 countries.
\end{ieeecsAcknowledgment}

\newpage


\ieeecsReferences{REFERENCES}

\bibliographystyle{ieeeCSBib}

\bibliography{refs}

\begin{thebibliography}{10}
\newcommand{\enquote}[1]{``#1''}
\providecommand{\url}[1]{\texttt{#1}}
\providecommand{\urlprefix}{}
\expandafter\ifx\csname urlstyle\endcsname\relax
  \providecommand{\doi}[1]{doi:\discretionary{}{}{}#1}\else
  \providecommand{\doi}{doi:\discretionary{}{}{}\begingroup
  \urlstyle{rm}\Url}\fi

\bibitem{siam-cse18}
U.~R\"uede et~al., \enquote{Research and Education in Computational Science and
  Engineering,} \emph{SIAM Review}, vol.~60, no.~3, 2018, pp. 707--754,
  \doi{10.1137/16M1096840}.

\bibitem{GroppHarrisonEtAl2016}
W.~Gropp, R.~Harrison et~al., \enquote{Future Directions for {NSF} Advanced
  Computing Infrastructure to Support {U.S.} Science and Engineering\ in
  2017-2020,} National Academies Press, 2016,
  \url{http://www.nap.edu/catalog/21886/}.

\bibitem{uk_survey}
S.~Hettrick, \enquote{2014 Software in Research Survey,} Feb. 2018,
  \doi{10.5281/zenodo.1183562}.

\bibitem{nangia_survey}
U.~Nangia and D.~S. Katz, \enquote{Track 1 Paper: Surveying the {US} National
  Postdoctoral Association Regarding Software Use and Training in Research,}
  \emph{Workshop on Sustainable Software for Science: Practice and Experiences
  (WSSSPE5.1)}, 2017, \doi{10.6084/m9.figshare.5328442.v3}.

\bibitem{nangia_nature}
U.~Nangia and D.~S. Katz, \enquote{Understanding Software in Research: Initial
  Results from Examining {N}ature and a Call for Collaboration,} \emph{13th
  IEEE International Conference on e-Science (e-Science)}, Oct 2017, pp.
  486--487, \doi{10.1109/eScience.2017.78}.

\bibitem{dia2}
\enquote{Deep Insights Anytime, Anywhere,} \url{http://www.dia2.org}, accessed:
  2018-08-31.

\bibitem{HerouxAllenEtAl2016}
M.~Heroux, G.~Allen et~al., \enquote{Computational Science and Engineering
  Software Sustainability and Productivity Challenges ({CSESSP}) Workshop
  Report,} 2016, {Networking and Information Technology Research and
  Development (NITRD) Program},
  \url{https://www.nitrd.gov/PUBS/CSESSPWorkshopReport.pdf}.

\bibitem{KeyesMcInnesWoodwardEtAl13}
D.~E. Keyes et~al., \enquote{Multiphysics Simulations: Challenges and
  Opportunities,} \emph{International Journal of High Performance Computing
  Applications}, vol.~27, no.~1, Feb 2013, pp. 4--83,
  \doi{10.1177/1094342012468181}.

\bibitem{SSI}
S.~Crouch et~al., \enquote{The Software Sustainability Institute: Changing
  Research Software Attitudes and Practices,} \emph{Computing in Science \&
  Engineering}, vol.~15, no.~6, Nov 2013, pp. 74--80, ISSN 1521-9615,
  \doi{10.1109/MCSE.2013.133}.

\bibitem{wenger}
E.~Wenger, \emph{Communities of Practice: Learning, Meaning, and Identity},
  Cambridge University Press, 1998, ISBN 978-0-521-66363-2.

\bibitem{rse_blog}
S.~Hettrick, \enquote{A Not-So-Brief History of Research Software Engineers,}
  \url{https://software.ac.uk/blog/2016-08-17-not-so-brief-history-research-software-engineers},
  2016, [Accessed: 2018-08-09].

\bibitem{urssi_cise}
J.~C. Carver et~al., \enquote{Conceptualization of a US Research Software
  Sustainability Institute (URSSI),} \emph{Computing in Science \&
  Engineering}, vol.~20, no.~3, May 2018, pp. 4--9, ISSN 1521-9615,
  \doi{10.1109/MCSE.2018.03221924}.

\bibitem{psip_tools}
M.~A. Heroux et~al., \enquote{Productivity and Sustainability Improvement
  Planning,}
  \url{https://bssw.io/resources/planning-for-better-software-psip-tools},
  2018, [Accessed: 2018-08-19].

\bibitem{bernholdt_ideas_outreach_blog2018}
D.~E. Bernholdt, \enquote{Think Locally, Act Globally: Outreach for Better
  Scientific Software,}
  \url{https://bssw.io/blog_posts/think-locally-act-globally-outreach-for-better-scientific-software},
  2018, [Accessed: 2018-08-19].

\bibitem{ideas_events_website}
\enquote{IDEAS events,} \url{https://ideas-productivity.org/events}, 2018,
  [Accessed: 2018-08-19].

\bibitem{bssw_contributing}
\enquote{Better Scientific Software: How to Contribute,}
  \url{https://bssw.io/contribute}, 2018, [Accessed: 2018-08-19].

\bibitem{SGCI-cise}
N.~Wilkins-Diehr and T.~D. Crawford, \enquote{NSF's Inaugural Software
  Institutes: The Science Gateways Community Institute and the Molecular
  Sciences Software Institute,} \emph{Computing in Science \& Engineering},
  vol.~20, no.~5, 2018, pp. 26--38, ISSN 1521-9615,
  \doi{10.1109/MCSE.2018.05329813}.

\bibitem{SGCI-road}
S.~Gesing et~al., \enquote{Science Gateways: The Long Road to the Birth of an
  Institute,} \emph{Proceedings of the 50th Hawaii International Conference on
  System Sciences (2017)}, Hawaii International Conference on System Sciences,
  2017, \doi{10.24251/hicss.2017.755}.

\bibitem{SGCI-2years}
N.~Wilkins-Diehr et~al., \enquote{The Science Gateways Community Institute at
  Two Years,} \emph{Proc. of Practice and Experience on Advanced Research
  Computing}, PEARC '18, ACM, 2018, pp. 53:1--53:8,
  \doi{10.1145/3219104.3219142}.

\bibitem{artaza}
H.~Artaza et~al., \enquote{Top 10 Metrics for Life Science Software Good
  Practices [version 1; ref: 2 approved],} \emph{F1000Research}, vol.~5, no.
  2000, 2016, \doi{10.12688/f1000research.9206.1}.

\bibitem{jimenez}
R.~Jim\'{e}nez et~al., \enquote{Four Simple Recommendations to Encourage Best
  Practices in Research Software [version 1; referees: 3 approved],}
  \emph{F1000Research}, vol.~6, no. 876, 2017,
  \doi{10.12688/f1000research.11407.1}.

\bibitem{ison}
J.~Ison et~al., \enquote{Tools and Data Services Registry: A Community Effort
  to Document Bioinformatics Resources,} \emph{Nucleic Acids Research},
  vol.~44, no.~D1, 2016, pp. D38--D47, \doi{10.1093/nar/gkv1116}.

\bibitem{cig-here}
W.~Bangerth et~al., \enquote{The role of Scientific Communities in Creating
  Reusable Software: Lessons from Geophysics,} \emph{Computing in Science \&
  Engineering}, 2018, \doi{10.1109/MCSE.2018.2883326}.

\bibitem{ropensci}
C.~Boettiger et~al., \enquote{Building Software, Building Community: Lessons
  from the rOpenSci Project,} \emph{Journal of Open Research Software}, vol.~3,
  no.~1, 2015, p.~e8, \doi{10.5334/jors.bu}.

\end{thebibliography}


\ieeecsAboutAuthor{ABOUT THE AUTHORS}

\begin{ieeecsAuthorBio}
\textbf{Daniel S. Katz} is Assistant Director for Scientific Software and Applications at the National Center for Supercomputing Applications (NCSA) and Research Associate Professor in Computer Science, Electrical and Computer Engineering, and School of Information Sciences (iSchool) at University of Illinois at Urbana-Champaign. His research interests include applications, algorithms, fault tolerance,  programming in parallel and distributed computing, citation and credit mechanisms and practices associated with software and data, organization and community practices for collaboration, and career paths for computing researchers. Katz received a PhD in Electrical Engineering from Northwestern University. Contact him at d.katz@ieee.org.
\end{ieeecsAuthorBio}

\begin{ieeecsAuthorBio}
\textbf{Lois Curfman McInnes} is a Senior Computational Scientist in the Mathematics and Computer Science Division of Argonne National Laboratory. Her research interests focus on high-performance computational science, with emphasis on scalable numerical libraries and collaboration toward productive and sustainable software ecosystems; she coordinates work on mathematical libraries in the U.S. Exascale Computing Project. McInnes received a Ph.D. in Applied Mathematics from the University of Virginia. Contact her at mcinnes@mcs.anl.gov.
\end{ieeecsAuthorBio}

\begin{ieeecsAuthorBio}
\textbf{David E. Bernholdt} is a Distinguished R\&D Staff Member and Group Leader at Oak Ridge National Laboratory. His research interests, broadly speaking, are related to making the development and use of software for HPC computational science and engineering more expressive, more productive, more sustainable, and better performing. Bernholdt received a Ph.D. in Chemistry from the University of Florida. Contact him at bernholdtde@ornl.gov.
\end{ieeecsAuthorBio}

\begin{ieeecsAuthorBio}
\textbf{Abigail Cabunoc Mayes} is a the Working Open Practice Lead at Mozilla. Her research interests include open source and prototyping. Cabunoc Mayes received a Bachelor of Computer Science, Honours Bioinformatics from the University of Waterloo. Contact her at abby@mozillafoundation.org.
\end{ieeecsAuthorBio}

\begin{ieeecsAuthorBio}
\textbf{Neil P. Chue Hong} is Director of the Software Sustainability Institute and based in the EPCC at the University of Edinburgh. His research interests include barriers and incentives in research software ecosystems and the role of software as a research object. Chue Hong received a Master of Physics in Computational Physics from the University of Edinburgh. Contact him at\\ N.ChueHong@software.ac.uk.
\end{ieeecsAuthorBio}

\begin{ieeecsAuthorBio}
\textbf{Jonah Duckles} is Director of Membership and Technology at The Carpentries. His research interests include ecological informatics, global agricultural production, graphical information systems, and remote sensing. Duckles received a Master of Science in Landscape Ecology from Purdue University. Contact him at jduckles@carpentries.org.
\end{ieeecsAuthorBio}

\begin{ieeecsAuthorBio}
\textbf{Sandra Gesing} is a Research Assistant Professor at University of Notre Dame. Her research interests include science gateways, computational workflows, and distributed computing. Gesing received a Ph.D. in Computer Science from University of T\"{u}bingen, Germany. Contact her at sandra.gesing@nd.edu.
\end{ieeecsAuthorBio}

\begin{ieeecsAuthorBio}
\textbf{Michael A. Heroux} is a Senior Scientist at Sandia National Laboratories, Scientist in Residence at St. John’s University, and Director of Software Technology for the U.S. Exascale Computing Project.  His research interests include all aspects of high-performance scientific computing and software.  Heroux received a Ph.D. in Applied Mathematics from Colorado State University.  Contact him at maherou@sandia.gov.
\end{ieeecsAuthorBio}

\begin{ieeecsAuthorBio}
\textbf{Simon Hettrick} is Deputy Director of the Software Sustainability Institute and co-Director of the Research Software Group at the University of Southampton. His work focuses on developing careers in software engineering within research organizations and studying the demographics of researchers who use software. Hettrick received a Ph.D. in laser physics from the University of Southampton. Contact him at s.hettrick@software.ac.uk.
\end{ieeecsAuthorBio}

\begin{ieeecsAuthorBio}
\textbf{Rafael C. Jimenez} is the Chief Data Architect at ELIXIR. His research interests include software development, computational workflows, data integration, data standards, data management, and data visualization. Jimenez received a M.Sc. in Computer Science from the University of KwaZulu-Natal. Contact him at rafael.jimenez@elixir-europe.org.
\end{ieeecsAuthorBio}

\begin{ieeecsAuthorBio}
\textbf{Marlon Pierce} is Director of the Science Gateways Research Center at Indiana University. His research interests include the application of distributed computing principals and modern software engineering practices to the development of open source middleware to support scientific research.  Pierce received a Ph.D. in Physics from Florida State University. Contact him at marpierc@iu.edu.
\end{ieeecsAuthorBio}

\begin{ieeecsAuthorBio}
\textbf{Belinda Weaver} is the Community and Communications Lead at The Carpentries. Her research interests include cloud computing, teaching coding, and digital skills for librarians. Weaver received a Graduate Certificate in Journalism from The University of Queensland. Contact her at bweaver@carpentries.org.
\end{ieeecsAuthorBio}

\begin{ieeecsAuthorBio}
\textbf{Nancy Wilkins-Diehr} is an Associate Director at the San Diego Supercomputer Center and principal investigator for the Science Gateways Community Institute. Her research interests include science gateways, high-end computing, and management of large distributed teams. Wilkins-Diehr received a M.S. in Aerospace Engineering from San Diego State University. Contact her at wilkinsn@sdsc.edu.
\end{ieeecsAuthorBio}


\end{document}